\documentclass[showpacs]{revtex4}
\usepackage{amsmath}
\usepackage{amssymb}
\usepackage{graphicx}

\begin{document}

\title{6D thick branes from interacting scalar fields}
\author{Vladimir Dzhunushaliev}
\email{dzhun@krsu.edu.kg}
\affiliation{Dept. Phys. and Microel.
Engineer., Kyrgyz-Russian Slavic University, Bishkek, Kievskaya Str.
44, 720021, Kyrgyz Republic}

\author{Vladimir Folomeev}
\email{vfolomeev@mail.ru}
\affiliation{Institute of Physics of NAS
KR, 265 a, Chui str., Bishkek, 720071,  Kyrgyz Republic}

\author{Douglas Singleton}
\email{dougs@csufresno.edu}
\author{Sergio Aguilar-Rudametkin}
\email{siarudametkin@hotmail.edu}
\affiliation{Physics Dept., CSU Fresno, Fresno, CA 93740-8031}

\begin{abstract}
A thick brane in six dimensions is constructed using two scalar fields.
The field equations for 6D gravity plus the scalar fields are solved numerically.
This thick brane solution shares some features with a previously
studied analytic solutions, but has the advantage that the energy-momentum tensor
which forms the thick brane comes from the scalar fields rather than being put in
by hand. Additionally the scalar fields which form the brane also provide a
universal, non-gravitational trapping mechanism for test fields of various spins.
\end{abstract}


\maketitle

\section{Introduction}

The brane world scenario, where our four dimensional spacetime is seen as lower dimensional
membrane embedded in a higher dimensional spacetime was proposed in \cite{arkani} \cite{gog} \cite{rs}
(see \cite{brane} for earlier work) as a means of addressing the hierarchy
problem. These brane models have also been used to investigate other open questions
in particle physics and cosmology such as the fermion generation puzzle \cite{fermi1} \cite{fermi2}
or the nature of dark energy \cite{de} and dark matter \cite{dm}.

In order to have an effective four dimensional spacetime in these brane world models
one must have a method of confining or trapping particles/fields to a four dimensional
spacetime. The localization of fields of various spins has been investigated
by various authors \cite{pom, bajc, local, ghsh}. In particular reference \cite{pom} showed
that it was not possible to trap spin 1 gauge bosons in the original 5D models of \cite{gog}
\cite{rs}. More specifically one could not trap spin 1 fields using gravity alone.
One had to introduce some other interaction to trap spin 1 fields. Further
in reference \cite{bajc} it was found that if one arranged the parameters of the
5D brane models of \cite{gog} \cite{rs} so that spin 0 and spin 2 fields were trapped then
spin $\frac{1}{2}$ would be repelled from the brane, and conversely if one arranged
the parameters so that spin $\frac{1}{2}$ fields were trapped then the spin 0 and
spin 2 field would be repelled.

In \cite{gosi1, gosi2} it was shown that by going from 5D to 6D it was possible to trap fields
of all spins (i.e. spins 0, $\frac{1}{2}$, 1, 2) to a 4D spacetime using only
gravity. The ``trapping" provided by the 6D solution of \cite{gosi2} is automatic since,
as we discuss shortly, the proper distance away from the brane at $r=0$ is
finite so any field can only be some finite distance away from the brane.
We briefly recall the features of this 6D brane solution that are relevant for the present paper.

First the 6D action was
\begin{equation}\label{6daction}
S = \int d^6x\sqrt {-g} \left[ \frac{M^4}{2}R +\Lambda + L_m
\right]~,
\end{equation}
where $M$, $R$, $\Lambda$ and $L_m$ are respectively the fundamental
scale, the scalar curvature, the cosmological constant and the
matter field Lagrangian. All of these physical quantities
refer to $6$-dimensional spacetime with the signature $(+ - - - - -)$.

Variation of the action \eqref{6daction} with respect to the
$6$-dimensional metric tensor $g_{AB}$ led to Einstein's
equations:
\begin{equation}\label{EinsteinEquation}
R_{AB} - \frac{1}{2}g_{AB} R = \frac{1}{M^4}\left( g_{AB} \Lambda
+ T_{AB} \right)~,
\end{equation}
where $R_{AB}$ and $T_{AB}$ are the Ricci and the energy-momentum
tensors respectively. Capital Latin indices run over $A, B,... =
0, 1, 2, 3, 5, 6 $.

The 4-dimensional Einstein equations were taken as
$R_{\mu\nu }^{(4)} - \frac{1}{2} \eta_{\mu\nu} R^{(4)} = 0$
i.e. the ordinary vacuum equations without no cosmological
constant. Greek indices $\alpha, \beta,... = 0, 1, 2, 3$ refer to four
dimensions. The source ansatz functions were
\begin{equation} \label{source}
T_{\mu\nu} =  - g_{\mu\nu} F(r), ~~~ T_{ij} = - g_{ij}K(r), ~~~
T_{i\mu} = 0 ~.
\end{equation}
Small Latin indices refer to the extra coordinates $i, j  = 5, 6$. Finally
the metric was taken to have the form
\begin{equation}\label{ansatzA}
ds^2  = \phi ^2 (r)\eta _{\mu \nu } dx^\mu
dx^\nu   - \lambda (r)(dr^2  + r^2d\theta ^2)~ ,
\end{equation}
The metric of ordinary 4-space,
$\eta_{\alpha \beta }$, has the signature $(+,-,-,-)$. The 4D and 2D
``warp" factor, ansatz functions $\phi (r)$ and $\lambda (r)$ depend only
on the extra radial coordinate, $r$. The Einstein field equations
\eqref{EinsteinEquation} in terms of the ansatz functions are \cite{gosi2}
\begin{eqnarray}\label{Einstein6a}
3 \frac{\phi^{\prime \prime}}{\phi} + 3 \frac{\phi ^{\prime}}{r
\phi} + 3 \frac{(\phi^{\prime})^2}{\phi ^2} +
\frac{1}{2}\frac{\lambda ^{\prime \prime}}{\lambda }
-\frac{1}{2}\frac{(\lambda^{\prime})^2}{\lambda^2} +
\frac{1}{2}\frac{\lambda^{\prime}}{r\lambda }
= \frac{\lambda }{M^4}[F(r) - \Lambda ] ~,  \\
\label{Einstein6b}
\frac{\phi^{\prime} \lambda^{\prime}}{\phi \lambda } + 2
\frac{\phi^{\prime}}{r\phi} + 3 \frac{(\phi^{\prime})^2}{\phi
^2} = \frac{\lambda }{2 M^4}[K(r) - \Lambda ] ~, \\
\label{Einstein6c}
2\frac{\phi^{\prime \prime}}{\phi} - \frac{\phi^{\prime} \lambda
^{\prime}}{\phi \lambda } + 3\frac{(\phi^{\prime})^2}{\phi^2} =
\frac{\lambda }{2 M^4}[K(r) - \Lambda ] ~,
\end{eqnarray}
where the prime $=\partial / \partial r$. These equations are for
the $\alpha \alpha$, $rr$, and $\theta \theta$ components
respectively. The three equations \eqref{Einstein6a} -- \eqref{Einstein6c}
are not all independent, and can be reduced to a set of
two equations for $\phi(r) , \lambda (r)$. We perform such
a reduction the following section.
An analytic solution to these equations was found
with the ansatz functions of the form
\begin{equation} \label{phi-lambda}
\phi (r) = \frac{c^2+ ar^2}{c^2+ r^2}~ \stackrel{r \rightarrow \infty}{\longrightarrow} a, \qquad 
\lambda (r) = \frac{c^4}{(c^2 + r^2)^2} \stackrel{r \rightarrow \infty}{\longrightarrow} \frac{c^4}{r^4}~.
\end{equation}
where $c, a$ are constant. The source functions are
\begin{equation} \label{FK}
F(r) = \frac{f_1}{2 \phi (r) ^2} +\frac{3 f_2}{4 \phi (r)}~, \stackrel{r \rightarrow \infty}{\longrightarrow}
\frac{f_1}{2 a ^2} +\frac{3 f_2}{4 a} ~~~~~ K(r) =
\frac{f_1}{\phi (r) ^2} +\frac{f_2}{\phi (r) }~ \stackrel{r \rightarrow \infty}{\longrightarrow} 
\frac{f_1}{a ^2} +\frac{f_2}{a },
\end{equation}
where the constants $f_1 = -\frac{3\Lambda}{5} a$ and $f_2=\frac{4
\Lambda}{5}(a+1)$ are determined by the 6D cosmological constant,
$\Lambda$, and the constant, $a$, from the 4D warp function $\phi
(r)$. 

A draw back of this solution is that the matter sources are put in by hand via the
ansatz functions $F(r)$ and $K(r)$ rather than being given by some realistic field source.
Also from $T_{\mu \nu}$ from \eqref{source} and $F (r)$ from \eqref{FK} (as well
as using the asymptotic values of $F(r)$ at $r=0$ and $r = \infty$) 
one sees that the energy density is negative on the brane and decreases to some 
negative, asymptotic, value at $r = \infty$. Other authors have used
such ``phantom" sources to construct brane world models \cite{koley}. 
In the present work we show that it is possible to replace the ``by-hand"
matter sources given by $F(r), K(r)$ by more realistic scalar field sources,
and still obtain the 4D and 2D ``warp" factors similar to those from \cite{gosi2}.
We also find that the energy density coming from the scalar fields has better
asymptotic behavior -- the energy density, while still negative on the brane,
approaches $0$ as $r \rightarrow \infty$. Finally we give a new, simple, non-gravitational
trapping mechanism which works for test fields of any spin. This new trapping mechanism
arises from one of the scalar fields which now replace the arbitrary source
ansatz functions, $F(r), K(r)$.

\section{6D brane from scalar fields}

We again start with 6D gravity and some matter source. The action is
\begin{equation}\label{6daction2}
S = \int d^6x\sqrt {-g} \left[ \frac{M^4}{2}R + L_m
\right]~,
\end{equation}
where we have apparently dropped the 6D cosmological constant and the
matter source is composed of two interacting scalar fields, $\chi (r)$ and
$\varphi (r)$ with the Lagrangian
\begin{equation}
\label{lagrangian}
    L_m =\frac{1}{2}\partial_A \varphi \partial^A
\varphi+\frac{1}{2}\partial_A \chi \partial^A \chi-V(\varphi,\chi)~,
\end{equation}
with the potential energy given by
\begin{equation}
\label{pot2}
    V(\varphi,\chi)=\frac{\Lambda_1}{4}(\varphi^2-m_1^2)^2+
    \frac{\Lambda_2}{4}(\chi^2-m_2^2)^2+\varphi^2 \chi^2-V_0.
\end{equation}
Where the constant $V_0$ is set as, $V_0 = (\Lambda _2 m_2 ^4 /4)$, 
and acts as a negative cosmological constant. In this paper the values
of $\Lambda _1, \Lambda _2$ were taken as $\Lambda _1 =0.1$ and
$\Lambda _2 =1.0$. For these values (and the associated $m_1, m_2$
which we find below) the potential \eqref{pot2} has 
two global minima at $\varphi=0, \chi = \pm m_2$
and two local minima at $\chi=0, \varphi = \pm m_1$. For all four of these
points one finds $\partial _\varphi V (\varphi , \chi) = 
\partial _\chi V (\varphi , \chi) = 0$. Also at these points
$V ( \pm m_1 , 0) = 0 > V (0 ,\pm m_2) = (\Lambda_1 m_1 ^4/4)-V_0$, 
which shows that the first two points are local minima. The fact that
$0 > V (0 ,\pm m_2)$ or $V_0 > \Lambda_1 m_1 ^4/4$ depends on the,
as yet undetermined, values of $m_1 , m_2$. Below we find $m_1 , m_2$
and they have values such that $V_0 > \Lambda_1 m_1 ^4/4$ which
makes the points $\varphi=0, \chi = \pm m_2$ global minima. Besides
two local and two global minima, four unstable saddle points exist:
$$
\varphi=\pm \sqrt{\frac{\Lambda_2}{2}}\sqrt{m_2^2-\frac{\Lambda_1
\Lambda_2 m_2^2-2 \Lambda_1 m_1^2}{\Lambda_1 \Lambda_2-4}}, \quad
\chi=\pm \sqrt{\frac{\Lambda_1
\Lambda_2 m_2^2-2 \Lambda_1 m_1^2}{\Lambda_1 \Lambda_2-4}}.
$$
The solutions which we find in this paper go to the local minima, 
$\chi=0, \varphi = \pm m_1$, and are thus not absolutely stable. However,
by adjusting parameters in the potential one can make the barrier between
the local and global minima large enough so that tunneling
between them is arbitrarily small, making the local minima meta-stable.

Other attempts to construct thick brane solutions
from scalar fields can be found in \cite{bronn} \cite{barbosa} \cite{rodriguez}
\cite{bazeia}. In some sense, with \eqref{lagrangian} \eqref{pot2} one is replacing
the two ansatz function, $F(r), K(r)$ of the previous solution by two scalar fields. The
two real scalar fields $\varphi, \chi$ depend only on the extra coordinate $r$; $m_1, m_2$
are the masses of these fields and $\Lambda_1, \Lambda_2$ are the self-coupling constants. The
effective, negative cosmological constant (i.e. the  $V_0$ term in the potential)
is the physical reason for the formation of the brane --
the attraction of the ordinary matter is balanced by the repulsion coming
from the negative cosmological constant. The potential in \eqref{pot2} was used in \cite{dzh06di} as
an approximate, effective description of a condensate of gauge field in SU(3) Yang-Mills theory
i.e. the scalar fields were taken as effective fields describing condensates of Yang-Mills
fields. In this view one can think of the brane as being formed from Yang-Mills
fields whose condensates are effectively represented by the scalar fields.
The general field equations for the scalar fields are given by
\begin{equation}
\label{scalar-eqn}
\frac{1}{\sqrt{-^6 \! g}}\frac{\partial}{\partial x^A}\left[\sqrt{-^6 \! g}\,\, g^{AB}
\frac{\partial (\varphi,\chi)}{\partial x^B}\right]=-\frac{\partial V}{\partial (\varphi,\chi)}.
\end{equation}
For the 6D metric given by \eqref{ansatzA} the $\varphi, \chi$ scalar field equations become
\begin{eqnarray}
\label{field-6d-1}
    \varphi^{\prime \prime}+\left(\frac{1}{r}+4\frac{\phi^{\prime}}{\phi}\right)
    \varphi^{\prime}=
    \lambda \varphi \left[ 2\chi^2 +\Lambda_1 (\varphi^2-m_1^2)\right]~, \\
    \chi^{\prime \prime}+\left(\frac{1}{r}+
    4\frac{\phi^{\prime}}{\phi}\right) \chi^{\prime}=
    \lambda \chi \left[ 2\varphi^2 +\Lambda_2 (\chi^2-m_2^2)\right]~.
\label{field-6d-2}
\end{eqnarray}
The Einstein field equations for the metric ansatz functions have the same left hand side as in
\eqref{Einstein6a}, but now the energy-momentum tensor on the right hand side comes from
the two scalar fields. The general form for the scalar field energy-momentum tensor is
\begin{equation}
\label{gen-em-scalar}
T_{AB} = \frac{\partial L_m}{\partial \varphi ^{,A}} \varphi _{,B}
+ \frac{\partial L_m}{\partial \chi ^{,A}} \chi _{,B} - g _{AB} L_m
\end{equation}
The specific components of the energy-momentum tensor are
\begin{eqnarray}
\label{em-scalar}
T_{\mu \nu} &=& g_{\mu \nu}   \left[ \frac{1}{2 \lambda} \left( \varphi^{\prime 2}+
        \chi^{\prime 2}\right) + V(\varphi,\chi) \right] \nonumber \\
T_{rr} &=& -\lambda \left[- \frac{1}{2 \lambda} \left( \varphi^{\prime 2}+
        \chi^{\prime 2}\right) + V(\varphi,\chi) \right] ~, \qquad
T_{\theta \theta} = - r^2 \lambda \left[ \frac{1}{2 \lambda} \left( \varphi^{\prime 2}+
        \chi^{\prime 2}\right) + V(\varphi,\chi) \right]~.
\end{eqnarray}
The Einstein field equations now become
\begin{eqnarray}\label{es6a}
3 \frac{\phi^{\prime \prime}}{\phi} + 3 \frac{\phi ^{\prime}}{r
\phi} + 3 \frac{(\phi^{\prime})^2}{\phi ^2} +
\frac{1}{2}\frac{\lambda ^{\prime \prime}}{\lambda }
-\frac{1}{2}\frac{(\lambda^{\prime})^2}{\lambda^2} +
\frac{1}{2}\frac{\lambda^{\prime}}{r\lambda }
= - \frac{\lambda }{M^4} \left[ \frac{1}{2 \lambda} \left( \varphi^{\prime 2}+
        \chi^{\prime 2}\right) + V(\varphi,\chi) \right] ~,  \\
\label{es6b}
\frac{\phi^{\prime} \lambda^{\prime}}{\phi \lambda } + 2
\frac{\phi^{\prime}}{r\phi} + 3 \frac{(\phi^{\prime})^2}{\phi
^2} = - \frac{\lambda }{2 M^4} \left[  - \frac{1}{2 \lambda} \left( \varphi^{\prime 2}+
        \chi^{\prime 2}\right) + V(\varphi,\chi) \right] ~, \\
\label{es6c}
2\frac{\phi^{\prime \prime}}{\phi} - \frac{\phi^{\prime} \lambda
^{\prime}}{\phi \lambda } + 3\frac{(\phi^{\prime})^2}{\phi^2} =
- \frac{\lambda }{2 M^4} \left[  \frac{1}{2 \lambda} \left( \varphi^{\prime 2}+
        \chi^{\prime 2}\right) + V(\varphi,\chi) \right] ~.
\end{eqnarray}
We can reduce these three equations to two by multiplying \eqref{es6c} by $\frac{3}{2}$
and subtracting from \eqref{es6a} to get a second order differential equation for $\lambda (r)$;
by subtracting \eqref{es6b} from \eqref{es6c} we get a second order differential equation
for $\phi (r)$.
\begin{eqnarray}
\label{Einstein-6d-1}
    \frac{\lambda^{\prime \prime}}{\lambda}-\left(\frac{\lambda^{\prime}}{\lambda}\right)^2-
    3 \left(\frac{\phi^{\prime}}{\phi}\right)^2+
    3\frac{\phi^{\prime}}{\phi}\frac{\lambda^{\prime}}{\lambda}+
    \frac{1}{r}
    \left( 6\frac{\phi^{\prime}}{\phi}+\frac{\lambda^{\prime}}{\lambda}\right )
    &=&
    -\frac{1}{2}\lambda \left[
        \frac{1}{2 \lambda} \left( \varphi^{\prime 2}+
        \chi^{\prime 2}\right) + V(\varphi,\chi)
    \right]~,
\\
    \frac{\phi^{\prime \prime}}{\phi}-\frac{\phi^{\prime}}{\phi}\frac{\lambda^{\prime}}{\lambda}-
    \frac{\phi^{\prime}}{r \phi}&=&
    -\frac{1}{4}\left( \varphi^{\prime 2}+ \chi^{\prime 2}\right)~.
\label{Einstein-6d-2}
\end{eqnarray}
The fundamental 6D gravity scale, $M$, has been adsorbed via the
rescaling: $r \rightarrow r/M^2$, $\varphi \rightarrow M^2 \varphi$,
$\chi \rightarrow M^2 \chi$, $m_{1,2} \rightarrow M^2 m_{1,2}$.

\begin{figure}[ht]
\begin{minipage}[t]{.5\linewidth}
  \begin{center}
  \fbox{
  \includegraphics[height=6cm,width=8.4cm]{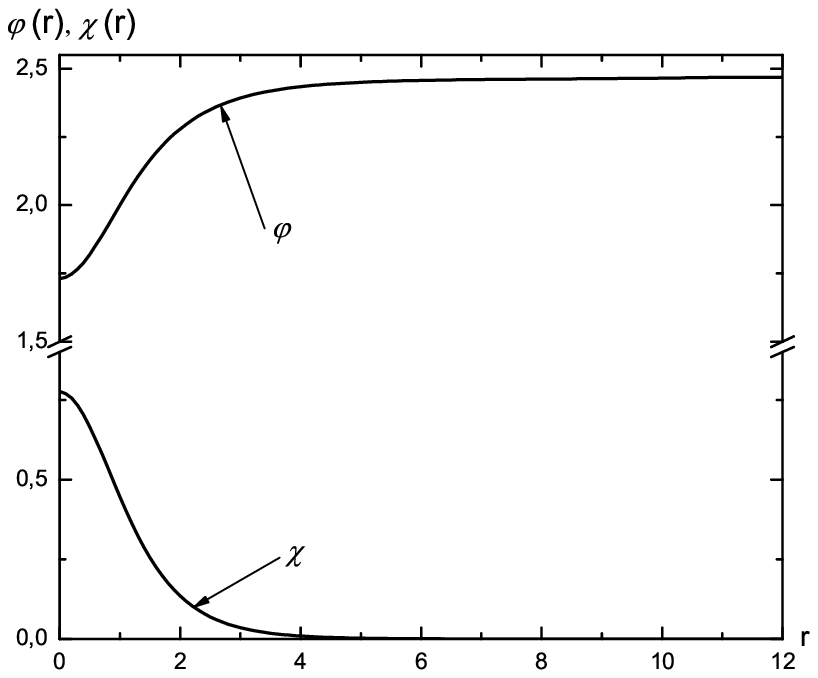}}
  \caption{The scalar fields $\varphi, \chi$ for the initial conditions \\
  given in \eqref{ini1}.}
    \label{phch1}
  \end{center}
\end{minipage}\hfill
\begin{minipage}[t]{.5\linewidth}
  \begin{center}
  \fbox{
  \includegraphics[height=6cm,width=8.4cm]{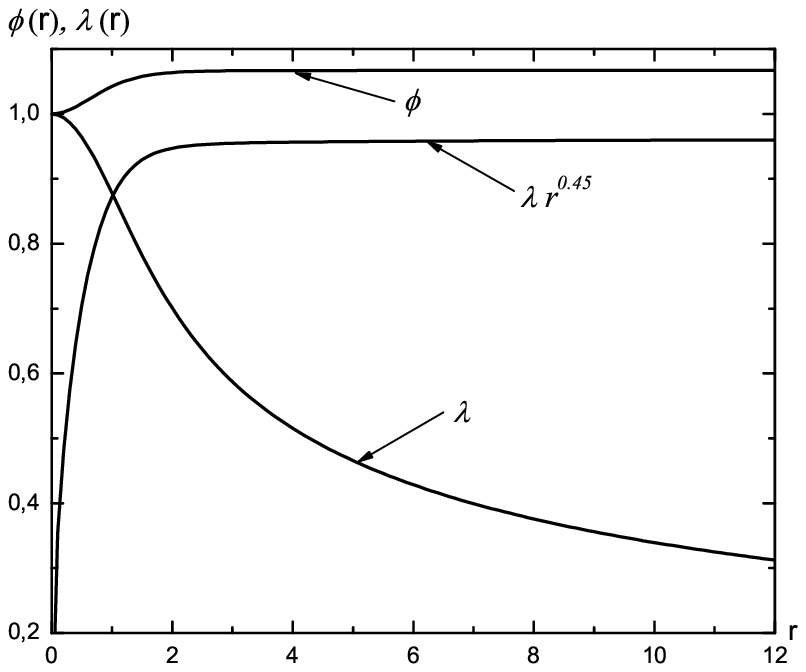}}
  \caption{The metric functions $\phi, \lambda$ for the initial conditions \\
      given in \eqref{ini1}. The middle curve indicates the power law fall
	  off for $\lambda \approx r^{-0.45}$}
  \label{met1}
  \end{center}
\end{minipage}\hfill
\end{figure}

We now show that the system of coupled, non-linear differential equations
\eqref{field-6d-1} -- \eqref{field-6d-2} and \eqref{Einstein-6d-1} -- \eqref{Einstein-6d-2} have solutions
which roughly share common features with the analytic brane solution given by \eqref{phi-lambda} -- \eqref{FK}.
Unlike the system of equations \eqref{Einstein6a} -- \eqref{Einstein6c} we were not able find an analytical
solution, but rather we solved the system numerically using the NDSolve routine from {\it Mathematica}.
As with the solutions in \eqref{phi-lambda} -- \eqref{FK} we require that the solution be 6D Minkowski on
the brane ($r=0$) so that $\phi (0) = 1$ and $\lambda (0) = 1$.
For initial condition we chose the ansatz functions at $r=0$ as
\begin{equation}
\label{ini1}
\varphi(0)=\sqrt{3} ~, ~ \varphi^\prime(0)=0, \qquad
\chi(0)=\sqrt{0.6} ~, ~  \chi^\prime(0)=0, \qquad
\phi(0)=1.0 ~ , ~  \phi^\prime(0)=0, \qquad
\lambda(0)=1.0 ~,~  \lambda^\prime(0)=0.
\end{equation}
Because of terms like $1/r$ we started
the NDSolve routine from $r=0.001$. Taking into account the vanishing of the first
derivatives of all the ansatz functions from \eqref{ini1}, all the ansatz functions
had an expansion of the form $f(r) = f(0) + f''(0) r^2 / 2$. The terms $f''(0)$ were determined
from one of the equations \eqref{field-6d-1} \eqref{field-6d-2} \eqref{es6a} or \eqref{es6c}. For
example $\phi (0.001) = \phi (0) + \frac{1}{2} (0.001)^2 \phi '' (0)$ and
$\phi ' (0.001) = (0.001) \phi '' (0)$ where $\phi '' (0) = - \frac{1}{4}
\phi (0) \lambda (0) V( \varphi (0) , \chi (0) )$ was obtained from \eqref{es6c} \eqref{pot2}
and \eqref{ini1}. The scalar field self couplings were taken as $\Lambda _1 =0.1$ and
$\Lambda _2 =1.0$. Once these initial conditions and scalar field
self couplings were set we searched for solutions which had good asymptotic
behavior i.e. we wanted the fields and metric ``warp" factor functions
to approach some constant, finite value as $r \rightarrow \infty$.
Such asymptotic conditions were only fulfilled for specific values of
$m_1, m_2$. For the initial conditions given in \eqref{ini1} and for
the chosen $\Lambda _1 , \Lambda _2$ we found that $m_1 \approx 2.462065$
and $m_2 \approx 3.0168291$ gave the desired asymptotic behavior. These
values of the masses, which gave the asymptotically well behave ansatz
functions, where found using the procedure outlined \cite{dzh-step}.
The numerical solutions for the scalar fields
and the metric ``warp" factor functions using the initial conditions
in \eqref{ini1} are shown in figures \eqref{phch1} and \eqref{met1}
respectively.
Note that since the ``warp" factors start as $\phi (0) = \lambda (0) = 1$ one
has 6D Minkowski spacetime on the brane, $r=0$. As $r \rightarrow \infty$ $\lambda (r) \rightarrow
0$ as $1/r ^\alpha$ while $\phi (0)$ approaches some constant asymptotic value $> 1$. These
features are generally similar to the ``warp" factors of the analytic solution given in \eqref{phi-lambda}.
However from figure \eqref{met1} one sees that for the present case that $\lambda (r)$ goes to
zero as $r ^{-0.45}$ which a much slower fall off as compared to the $\lambda (r)$ given 
in \eqref{phi-lambda}. The scalar fields approach the asymptotic values
$\varphi (\infty) = m_1$ and $\chi (\infty ) = 0$. Looking at the potential, $V(\varphi , \chi)$,
from \eqref{pot2} one sees that $V (\varphi (\infty) = m_1 , \chi (\infty) =0 ) = 0 >
V (\varphi (\infty) = 0 , \chi (\infty)= m_2 ) = -V_0$. Thus the solution is not absolutely stable since
it sits at a local rather than global minimum. However, by adjusting parameters in the potential
$V(\varphi , \chi)$ one can make the barrier between
the local and global minimum large. Thus tunneling between the local and global
minimum will be unlikely making this state
with $\varphi (\infty) = m_1$ and $\chi (\infty ) = 0$ effectively stable over 
long time scale e.g. long compared to the age of the Universe.

\begin{figure}[ht]
\begin{center}
\fbox{
  \includegraphics[height=6cm,width=8.4cm]{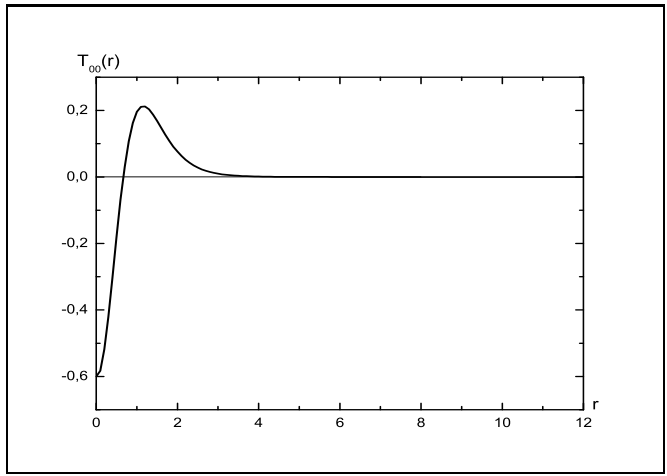}}
 \caption{The energy density $T_{00} (r)$ from \eqref{em-scalar}. $T_{00} (r) <0$ for $r <1$}
\label{energ1}
\end{center}
\end{figure}

Once the scalar fields, $\varphi (r), \chi (r)$, are known it is possible to calculate
their energy density. Plugging the numerical solutions for the scalar fields into \eqref{em-scalar}
the energy density is given by figure \eqref{energ1}. Both the energy density of the scalar
fields system, given in  figure \eqref{energ1}, and the energy density of the analytical solution,
given by equations \eqref{source} \eqref{FK}, approached a constant value as $r \rightarrow \infty$. However for 
the scalar fields case the asymptotic value was zero, while for the analytic solution it was a negative
constant. Both solutions have some range of $r$, for which $T_{00}< 0$ -- for the scalar fields
this region is only near the brane (see figure \eqref{energ1}) while for the analytic solution 
$T_{00} <0$ for all $r$. In the scalar fields case this negative energy density 
may provide a physical explanation for the formation of the brane:
the negative energy density provides a repulsive force which can balance the
usual attraction due to gravity.

We now analyze the asymptotic forms for the solutions to the scalar field equation,
\eqref{field-6d-1} and \eqref{field-6d-2}, and the Einstein field equations,
\eqref{es6a} - \eqref{es6c}. The asymptotic forms of the ansatz functions can
be written as:
\begin{equation}
\label{asymptotic}
    \varphi\approx m_1-\delta \varphi (r), \quad \chi\approx \delta \chi (r),
    \quad \phi\approx \phi_{\infty},
\quad \lambda \approx \frac{\gamma}{r^\alpha}.
\end{equation}
Where $\delta \varphi (r) , \delta \chi (r) \ll 1$ as $r \rightarrow \infty$, and $\gamma$ is some
dimensionful constant. The subscript ``$\infty$'' indicates the asymptotic value of $\phi (r)$.
Note in particular that the 2D metric ansatz function, $\lambda (r)$ has a non-trivial
behavior as $r^{-\alpha}$. This comes about since this form gives zero when used in the
three terms involving $\lambda$ and its derivatives on the left hand side of \eqref{es6a}.
The three terms involving $\phi$ and its derivatives on the left hand side of
\eqref{es6a} yields to a similar analysis, but the minus sign difference in this case
results in the exponent being zero so that $\phi \approx r^0 =$constant. From the numerical
analysis -- see figure \eqref{met1} -- the exponent for the 2D metric ansatz function is
$\alpha = 0.45$ so that $\lambda \approx \gamma / r^{-0.45}$.
One can see -- ignoring terms that are second order in small quantities like $\varphi ' (r) ,
\chi, \chi ' (r)$ -- that the Einstein field equations given by \eqref{es6a} - \eqref{es6c} are satisfied the
asymptotic forms in \eqref{asymptotic}. The right hand sides of \eqref{es6a} - \eqref{es6c}
goes to zero since both $\varphi^{\prime 2} , \chi^{\prime 2} \rightarrow 0$ and
$V(\varphi,\chi) \rightarrow 0$ as $r \rightarrow \infty$ for the asymptotic forms given
in \eqref{asymptotic}. Furthermore one can see the left hand sides also go to zero for the
asymptotic forms for $\phi (r) , \lambda (r)$ given in \eqref{asymptotic}.

Turning to the asymptotic form for the scalar field equations --
\eqref{field-6d-1} and \eqref{field-6d-2} -- we find that for the
asymptotic forms from \eqref{asymptotic} these equations become
\begin{equation}
\label{asympt1}
    \delta \varphi^{\prime \prime}+\frac{1}{r} \delta \varphi^{\prime}=
    \frac{2\gamma \Lambda_1 m_1^2}{r^\alpha} \delta \varphi ~, \qquad
    \delta \chi^{\prime \prime}+\frac{1}{r} \delta \chi^{\prime}=
    \frac{\gamma(2 m_1^2-\Lambda_2 m_2^2)}{r^\alpha} \delta \chi,
\end{equation}
These equations have the solutions:
\begin{eqnarray}
    \delta \varphi &=&
    C_{\varphi} K_0 \left(
        \frac{2\sqrt{2 \gamma \Lambda_1 m_1^2}}{2-\alpha} r^{1-\alpha/2}
    \right)
    ~ \approx ~ C_{\varphi}
    \frac{\exp{\left(-\frac{2\sqrt{2 \gamma \Lambda_1 m_1^2}}{2-\alpha} r^{1-\alpha/2}\right)}}
    {\sqrt{r^{1-\alpha/2}}},
    \label{sol1} \\
    \delta \chi &=& C_{\chi} K_0 \left(
        \frac{2\sqrt{ \gamma (2 m_1^2-\Lambda_2 m_2^2)}}{2-\alpha} r^{1-\alpha/2}
    \right)
    ~ \approx ~ C_{\chi} \frac{\exp{\left(
        -\frac{2\sqrt{2 \gamma (2 m_1^2-\Lambda_2 m_2^2)}}{2-\alpha} r^{1-\alpha/2}
    \right)}}
    {\sqrt{r^{1-\alpha/2}}},
\label{sol2}
\end{eqnarray}
where $K_0 (C r)$ are the zeroth order modified Bessel function of second kind, which 
asymptotically decay as $e^{-C r} / \sqrt{r}$. This analysis indicates that asymptotically the
ansatz functions are well behaved, and this analytic, asymptotic analysis matches the
numerical solutions shown in figures \eqref{phch1} \eqref{met1}. 

From \eqref{sol1}-\eqref{sol2}, one can see that it is necessary to
take $\alpha<2$ in order that the scalar fields decay as $r \rightarrow
\infty$. The numerical analysis in our case gives $\alpha \approx
0.45$. Thus as indicated in figure \eqref{phch1} and confirmed by the asymptotic forms
with $\alpha \approx 0.45$ we do have decaying solutions for the scalar
fields (in the case of the $\varphi$ field it is ``decaying" to a non-zero constant).

One final, and important, difference between the thick brane solution of the present section and the
analytic solution of the previous section is that the proper distance for the present 
solution is infinite. This can be seen directly by calculating the proper length 
\begin{equation}
    l = \int \limits_0^{r_0} \sqrt{\lambda (r)} dr \approx \int \limits_0^{r_0} \sqrt{\frac{\gamma}{r^{0.45}}} dr
    \stackrel{r_0 \rightarrow \infty}{\longrightarrow} \infty  ~,
\label{prop}
\end{equation}
which is infinite. For the analytic solution of the previous section $\lambda (r) \rightarrow c^4 /r^4$
from \eqref{phi-lambda}. Using this asymptotic form in \eqref{prop} gives a finite proper length.

Because the proper length away from the brane
is infinite rather than finite -- as was the case for the analytic solution of
\eqref{phi-lambda} -- we need to revisit the question of whether
fields of various spins are trapped or not. For the analytic solution of \eqref{phi-lambda} 
this question was investigated in detail in \cite{gosi1} \cite{gosi2} and not unsurprisingly 
all the various spin fields where ``trapped" exactly because the proper 
length was finite. The criteria for trapping taken in 
\cite{gosi1} \cite{gosi2} was that the integration of the
action for some test field of spin-$0, \frac{1}{2}, 1, 2$ over the extra dimensions
should be finite. This is the criteria for trapping we take in the present work.
One can also take as the criteria for trapping that total field energy per unit
3-volume of the brane (i.e. the integral of the energy density, $T_0 ^{~0}$ over
the extra spatial dimensions) be finite \cite{bronn}. In the present
work these two criteria give the same result in regard to whether or
not the field is trapped for all cases we consider. 

We now investigate in detail the trapping of a spin-$0$ field. The spin-$1$
and spin-$2$ cases are identical to the spin-$0$ case. The spin-$1/2$ case
can be worked out along the same lines as the other fields using the
set-up given in \cite{gosi1}. The details are somewhat different than for the integer spin
fields but the final conclusion -- that the spin-$1/2$ and all other spin fields are not 
trapped -- is the same. We begin by considering a real scalar field $S(x^A)$ with an 
action given by
\begin{equation}
\label{test0}
S_0 = - \frac{1}{2} \int d^6x g ^{MN} \partial _M S \partial _N S
\end{equation}
The field equations for $S(x^A)$ resulting from \eqref{test0} are similar to those
for $\varphi, \chi$ \eqref{scalar-eqn} but without a potential term
\begin{equation}
\label{test0a}
\frac{1}{\sqrt{-^6 g}} \partial _A [ \sqrt{- ^6 g} g^{MN} \partial _N S(x^A)] =0.
\end{equation}
Making a separation of variables as $S(x^A) = \Sigma (x^\mu ) s(r)$ one
finds \cite{oda} that \eqref{test0a} is solved by $s(r) = C_0 = const.$
and $\eta ^{\mu \nu} \partial _\mu \partial _\nu \Sigma =0$ i.e. the 4D
scalar wave function is massless. Inserting this solution into the action
\eqref{test0} one obtains
\begin{equation}
\label{scalar-action}
S_0 = - \pi \int_0 ^\infty dr ~ \left[ r \phi ^2 (r) \lambda (r) \right] ~
\int d^4 x \sqrt {- \eta} \left[\eta ^{\mu \nu} \partial _\mu  S \partial _\nu S \right] ~
\stackrel{r \rightarrow \infty}{\longrightarrow} -  \pi C_0^2 \phi _\infty ^2 \gamma
\int_0 ^\infty ~ r^{0.55} dr ~ \int d^4 x \sqrt {- \eta} \left[\eta ^{\mu \nu} \partial _\mu  \Sigma \partial _\nu \Sigma \right] ~.
\end{equation}
where we have used the asymptotic forms $\phi (\infty) = \phi _\infty$ and $\lambda (\infty) =\gamma / r^{0.45}$.
The integral over $r$ in the last line of \eqref{scalar-action} is divergent at large $r$. Thus the test scalar
field is not trapped by the metric given by $\phi (r)$ and $\lambda (r)$. The cases of spin-$1$ and spin-$2$
fields also yield integral which diverge as $\int_0 ^\infty ~ r^{0.55} dr$. The spin-$1/2$
case is different in detail but also yields a divergent integral which goes as $\int_0 ^\infty ~ r^{-0.33} dr$. 
Thus none of the fields are trapped by the metric alone. In the next section we show that the
exponential form for the scalar field $\chi (r)$ can give a universal trapping mechanism. 

Before moving on to the trapping mechanism we make a few comments about the geometric 
character of the metric given by figure \eqref{met1} and by the asymptotic expressions
\eqref{asymptotic}. The metric solution is a co-dimension 2 object which is
asymptotically flat as $r \rightarrow \infty$ and has a deficit angle 
of $2 \pi (1 - \sqrt{1-\frac{\alpha}{2}})$.
To see this we change the asymptotic form of the metric (given by inserting \eqref{asymptotic} in
\eqref{ansatzA}) via the transformation $R=r^{1-\frac{\alpha}{2}}$. With this the
2D part of the metric becomes 
$$
ds_{2D}^2 \rightarrow \frac{\gamma}{1 - \frac{\alpha}{2}}
\left(dR^2 + \left(1-\frac{\alpha}{2} \right) R^2 d \theta^2 \right) 
$$
For this 2D metric one defines a new angle ${\bar \theta} = \theta \sqrt{1-\frac{\alpha}{2}}$ 
so that $0 \le {\bar \theta} \le 2 \pi \sqrt{1-\frac{\alpha}{2}}$ which gives a 
deficit angle of $2 \pi (1 - \sqrt{1-\frac{\alpha}{2}})$. Since $\alpha = 0.45$
this is a deficit angle rather than a proficit angle. 
Recently there has been interest in such brane world models with deficit 
\cite{navarro} and proficit \cite{gosijhep} angles to investigate open 
questions such as the cosmological constant and the fermion generation puzzle.

\section{Trapping mechanism}

All the test fields studied in the previous subsections have divergent actions coming from the
integration over $r$. Thus none of these fields are trapped to the brane at $r=0$ and
as it stands the model is not viable. However, the form of the scalar field, $\chi (r)$, provides
a mechanism for trapping all the fields studied in the previous subsections. Note that all the spin
fields have $r$ integrals that diverge as some power ($r^{1.55}$ for spin 0, spin 1 and spin 2;
$r^{0.67}$ for spin $\frac{1}{2}$). On the other hand from \eqref{sol2} one sees that $\chi (r)$
decreases exponentially. This exponential decrease depends on several things: (i) $\gamma >0$
(ii) $m_1 ^2 > \frac{\Lambda _2 m_2 ^2}{2}$ (iii) $\alpha <2$. If (i) or (ii) is not satisfied
then $\chi (r)$ is oscillatory; if (ii) is not satisfied $\chi (r)$ grows exponentially. For our solution
all three conditions are met so $\chi (r)$  decreases exponentially. Since the decrease of
$\chi (r)$ is exponential while the divergence of the $r$ integration for each of the fields
is some power law $r^b$ with $b<1$  one can get the $r$ integration to converge by multiplying
the Lagrangian density of each spin field by some positive power of $\chi (r)$ as
\begin{equation}
\label{trapping-chi}
S_s = \int d^6 x ~ \chi ^n (r) ~ L _s ~,
\end{equation}
where $n >0$ and $s = 0, \frac{1}{2} , 1, 2$. Redoing the analysis of the previous subsections
it is easy to see that this procedure will make each of the $r$ integrations converge thus
giving a trapping of the test field.

An even more stringent and less {\it ad hoc} trapping mechanism can be obtained
by introducing a dilaton-like exponential coupling as suggested in the seminal paper
\cite{callan} \cite{russo}. As in these works we can introduce an exponential, dilaton-like
coupling between $\chi (r)$ and the various spin fields in the following way
\begin{equation}
\label{dilaton}
S_s = \int d^6 x ~ (1 - e^{-2 \chi (r)} )~ L _s
\end{equation}
Since $\chi (r)$ exponentially goes to zero the factor $1 - e^{-2 \chi (r)}$
will make the $r$ integration for each of the various spin test fields strongly
convergent. In each case (spin 0, $\frac{1}{2}$, 1, 2) we find the fields are exponentially
trapped by the behavior of $\chi (r)$ to some small region near the brane at $r=0$.

The non-gravitational, trapping mechanism suggested by \eqref{trapping-chi} or \eqref{dilaton}
is simple and universal. It is made possible by the asymptotic behavior of the scalar field $\chi (r)$ which
plays a dual role of forming the brane and trapping test fields of all spin to the brane.
This mechanism is may be compared in some respects to confinement in quantum chromodynamics or
to the confinement of magnetic charges inside a superconductor. Particularly in the
magnetic charges inside a superconductor
example, the scalar field condensate of Cooper pairs plays a crucial role.

\section{Conclusions}

We have constructed a thick brane solution in 6D spacetime using two self-interacting,
and mutually interacting scalar fields. This thick brane solution had the same general
characteristics as the analytic solution given \eqref{phi-lambda} \eqref{FK}: the warp
factors, $\phi (r), \lambda (r)$ were equal to $1$ at $r=0$ so that on the brane
one had a 6D Minkowski spacetime. As $r \rightarrow \infty$ the warp factor functions
approached constant asymptotic values $\phi (\infty) >1$ and $\lambda (\infty) = 0$.
The energy densities were also similar -- both were negative on the brane and approached 
some different, asymptotic value as $r \rightarrow \infty$. Two important advantages of the
present solution is that the asymptotic value of the energy density in the bulk
was zero rather than negative. In both cases the negative energy density may
provide a physical explanation for the formation of the brane -- the
repulsion from the negative energy density can balance the attraction due
to gravity. Another advantage of the present solution
is that the energy-momentum tensor for the previous analytic
solution was ``fixed by hand" in order to give the warp factors in \eqref{phi-lambda}
which provided the universal gravitational trapping of particles/fields of all
spins. In the present case the energy-momentum tensor comes from a more
realistic source i.e. two scalar fields. In essence the two ansatz functions
$F(r), K(r)$ from \eqref{FK} have been replaced by the two scalar fields
$\varphi (r) , \chi (r)$.

Because the 2D warp factor, $\lambda (r)$, goes to zero as
$r \rightarrow \infty$ according to a power law \eqref{asymptotic} with $\alpha<2$,
the proper distance away from the brane is
infinite, rather than finite as in the case of the analytic solution
in \cite{gosi2}. Thus we were forced to introduce a new, simple, non--gravitational trapping
mechanism for test fields of various spins moving in the background
\eqref{ansatzA} with $\phi (r), \lambda (r)$ given by figure \eqref{met1}.
The mechanism involves multiplying the test field Lagrange density
for a test field of spin $s$ by some positive power of the scalar field $\chi (r)$ as
in \eqref{trapping-chi} \eqref{dilaton}. Thus $\chi (r)$ not only forms the brane near $r=0$
with an energy density peaked near $r=0$ (see figure \eqref{energ1}) and going
to zero at $r=\infty$, but also is responsible for the trapping of
test fields of various spins to the brane.

The brane solution found here is not absolutely stable since it settles into
one of the local minima at $\varphi (\infty) = m_1$ and $\chi (\infty ) = 0$ rather
than one of the global minima at $\varphi (\infty) = 0$ and $\chi (\infty ) = \pm m_2$.
However, the barrier between the local and global minima can be made large by adjusting the 
parameters the scalar potential \eqref{pot2}. For example, one could chose different
values of $\Lambda_1 , \Lambda _2$. In this way the decay probability of the present solution from the
local minimum to the global minimum via tunneling could be made small. This would make 
the brane solution presented here effectively stable over long time scales. One open question 
not studied in the present paper is the stability of this solution in the Lyapunov sense
(see for example \cite{rubakov}). We leave this question for future study. 

\section{Acknowledgment}

V.D. acknowledges D. Singleton for the invitation to do research at Fresno
State University and the support of a CSU Fresno Provost Award Grant. D.S. acknowledges
the support of a 2007 Summer Professional Development Grant from the CSM of CSU Fresno.

\end{document}